\documentclass[prc,twocolumn,showpacs,superscriptaddress,floatfix, aps, prc]{revtex4-1}
\usepackage{graphicx}
\usepackage{dcolumn}
\usepackage{bm}
\usepackage{mwe,tikz}
\usepackage[percent]{overpic}
\usepackage{multirow}
\usepackage[export]{adjustbox}

\begin{document}


\title{Spectroscopy of nuclei around $^{100}$Sn populated via two-neutron knockout reactions}

\author{A.$~$Corsi}
\email{acorsi@cea.fr}
\affiliation{IRFU, CEA, Universit\'e Paris-Saclay, 91191 Gif-sur-Yvette, France}

\author{A.$~$Obertelli}
\affiliation{IRFU, CEA, Universit\'e Paris-Saclay, 91191 Gif-sur-Yvette, France}
\affiliation{RIKEN Nishina Center, 2-1 Hirosawa, Wako, Saitama 351-0198, Japan}
\affiliation{Institut fur Kernphysik, TU Darmstadt, D-64289 Darmstadt, Germany}

\author{P.$~$Doornenbal}
\affiliation{RIKEN Nishina Center, 2-1 Hirosawa, Wako, Saitama 351-0198, Japan}


\author{F.$~$Nowacki}
\affiliation{IPHC, CNRS/IN2P3, Universit\'e de Strasbourg, F-67037 Strasbourg, France}


\author{H.$~$Sagawa}
\affiliation{University of Aizu, Aizu, Japan}

\author{Y.$~$Tanimura}
\affiliation{Institut de Physique Nucl\'eaire, Universit\'e Paris-Saclay, 91000 Orsay, France}
\affiliation{Department of Physics, Tohoku University, Sendai 980-8578, Japan}

\author{N.$~$Aoi}
\affiliation{RCNP Osaka University, Ibaraki, Osaka 567-0047, Japan}

\author{H.$~$Baba}
\affiliation{RIKEN Nishina Center, 2-1 Hirosawa, Wako, Saitama 351-0198, Japan}

\author{P.$~$Bednarczyk}
\affiliation{The Niewodniczanski Institute of Nuclear Physics, Krakow, Poland}

\author{S.$~$Boissinot}
\affiliation{IRFU, CEA, Universit\'e Paris-Saclay, 91191 Gif-sur-Yvette, France}

\author{M.$~$Ciemala}
\affiliation{The Niewodniczanski Institute of Nuclear Physics, Krakow, Poland}

\author{A.$~$Gillibert}
\affiliation{IRFU, CEA, Universit\'e Paris-Saclay, 91191 Gif-sur-Yvette, France}

\author{T.$~$Isobe}
\affiliation{RIKEN Nishina Center, 2-1 Hirosawa, Wako, Saitama 351-0198, Japan}

\author{A.$~$Jungclaus}
\affiliation{Instituto de Estructura de la Materia, CSIC, E-28006 Madrid, Spain}

\author{V.$~$Lapoux}
\affiliation{IRFU, CEA, Universit\'e Paris-Saclay, 91191 Gif-sur-Yvette, France}

\author{J.$~$Lee}
\affiliation{RIKEN Nishina Center, 2-1 Hirosawa, Wako, Saitama 351-0198, Japan}
\affiliation{Department of Physics, The University of Hong Kong, Pokfulam Road, Hong Kong, China}

\author{K.$~$Matsui}
\affiliation{RIKEN Nishina Center, 2-1 Hirosawa, Wako, Saitama 351-0198, Japan}

\author{M.$~$Matsushita}
\affiliation{CNS, The University of Tokyo, RIKEN Campus, Wako, Saitama 351-0198, Japan}

\author{T.$~$Motobayashi}
\affiliation{RIKEN Nishina Center, 2-1 Hirosawa, Wako, Saitama 351-0198, Japan}

\author{D.$~$Nishimura}
\affiliation{Department of Physics, Tokyo City University, Setagaya-ku, Tokyo 158-8557, Japan}

\author{S.$~$Ota}
\affiliation{CNS, The University of Tokyo, RIKEN Campus, Wako, Saitama 351-0198, Japan}

\author{E.$~$Pollacco}
\affiliation{IRFU, CEA, Universit\'e Paris-Saclay, 91191 Gif-sur-Yvette, France}

\author{H.$~$Sakurai}
\affiliation{RIKEN Nishina Center, 2-1 Hirosawa, Wako, Saitama 351-0198, Japan}

\author{C.$~$Santamaria}
\affiliation{IRFU, CEA, Universit\'e Paris-Saclay, 91191 Gif-sur-Yvette, France}
\affiliation{National Superconducting Cyclotron Laboratory and Department of Physics and Astronomy, Michigan State University, East Lansing, Michigan 48824, USA}

\author{Y.$~$Shiga}
\affiliation{Department of Physics, Rikkyo University, Toshima, Tokyo 172-8501, Japan}

\author{D.$~$Sohler}
\affiliation{Institute for Nuclear Research, Debrecen, Hungary}

\author{D.$~$Steppenbeck}
\affiliation{CNS, The University of Tokyo, RIKEN Campus, Wako, Saitama 351-0198, Japan}

\author{S.$~$Takeuchi}
\affiliation{RIKEN Nishina Center, 2-1 Hirosawa, Wako, Saitama 351-0198, Japan}
\affiliation{Department of Physics, Tokyo Institute of Technology, 2-12-1 O-Okayama, Meguro, Tokyo 152-8551, Japan}

\author{R.$~$Taniuchi}
\affiliation{Department of Physics, University of Tokyo, Bunkyo, Tokyo 113-0033, Japan}

\author{H.$~$Wang}
\affiliation{RIKEN Nishina Center, 2-1 Hirosawa, Wako, Saitama 351-0198, Japan}

\date{\today}

\begin{abstract}
We report on the in-beam gamma spectroscopy of $^{102}$Sn and $^{100}$Cd produced via two-neutron removal from carbon and CH$_2$ targets at about 150 MeV/nucleon beam energy. New transitions assigned to the decay of a second 2$^+$ excited state at 2470(60) keV in $^{102}$Sn were observed. Two-neutron removal cross sections from $^{104}$Sn and $^{102}$Cd have been extracted. The enhanced cross section to the 2$^+_2$ in $^{102}$Sn populated via the $(p,p2n)$ reaction is traced back to an increase of shell-model structure overlaps, consistent with the hypothesis that the proton-induced two-deeply-bound-nucleon removal mechanism is of direct nature. 
\end{abstract}

\maketitle


\section{\label{sec:level1} Introduction}


Tin nuclei, with 42 known isotopes and a proton closed shell, are an ideal testing ground to study the evolution of shell structure and pairing correlations, which are expected to be the dominant many-body correlations in these Z=50 nuclei \cite{pot11}. A transition from superfluid nuclei at mid-shell to spherical nuclei is expected approaching the neutron shell closures at N=50 and 82,  where the seniority scheme can be adopted to describe the energy spectra and transition strengths. This phase transition is confirmed by the evolution of B(E2) strengths, showing large constant B(E2) values in the mid-shell region and smaller ones going towards the N=50 and 82 isotopes. In the seniority scheme, transition probabilities are proportional to the product of the number of particle pairs and hole pairs in a given orbital and therefore are expected to be maximum at mid shell and to decrease toward shell closures. Recently, light tin isotopes have attracted significant interest because they deviate from this scheme when approaching $^{100}$Sn \cite{gua13, bad13, dor13, cor15}. In Ref. \cite{cor15}, the study of inclusive inelastic $(p,p')$ cross sections supports the interpretation that the proton collectivity in light Tin isotopes is driven by the neutrons, as predicted by QRPA calculations. Such an onset of collectivity is not observed in light Cd isotopes, that follow the expected pattern with collectivity decreasing towards $^{100}$Cd \cite{eks09}, showing the specificity of the Tin isotopic chain. \\
Two-neutron removal cross sections might also show the signature of the transition from superfluid to non-superfluid nuclei, as will be discussed in this paper.\\
The filling of a (sub-)shell is expected to quench the pairing correlations of the ground state and give rise to appreciable population of low-lying excited states via two-neutron transfer \cite{yos62, BHR}. 
As an example, a decrease of the $(p,t)$ transfer cross section to the ground state of $^{114}$Sn, with N=64 corresponding to the filling of the d$_{5/2}$-g$_{7/2}$ shell and behaving as sub-shell closure, was observed \cite{bje68}. More recently, two-neutron $(p,t)$ transfer reactions have been systematically used to extract cross sections to ground and excited states in Tin isotopes from $^{110}$Sn to $^{122}$Sn \cite{gua99,gua04,gua06,gua08,gua11,gua12}. Absolute values of the $(p,t)$ cross sections to $^{110-122}$Sn could be well reproduced with a DWBA calculation with two-nucleon transfer amplitudes calculated with a pairing potential adjusted to experimental odd-even mass differences \cite{pot11}, thus confirming that pairing plays a key role in the collectivity of Tin isotopes and drives the two-nucleon transfer mechanism. A drop of the ratio of the two-neutron transfer cross sections to 0$^+_1$ over higher 0$^+_i$ states, $\sigma_{0^+_1}$/$\sigma_{0^+_{i>1}}$, appears at N=64, 66, 70, 72, with a complex pattern that cannot be explained only by the N=64 shell closure. \\
No spectroscopic information is available on $^{100}$Sn, while for its even-even neighbors $^{100}$Cd and $^{102}$Sn the available spectroscopic information comes mainly from fusion-evaporation experiments, which populate preferentially high-spin states decaying to the yrast band. The spectroscopy of $^{102}$Sn is known for yrast states up to the 6$^+$ isomer at 2017 keV \cite{lip98}. The spectroscopy of $^{100}$Cd is known for yrast states up to 20$\hbar$ in angular momentum and 10 MeV in excitation energy \cite{gor94, cla00}. \\
For exotic unstable nuclei, the extraction of spectroscopic information from two-nucleon transfer reactions is experimentally impracticable due to the limited beam intensity available at energies relevant for nucleon-transfer reactions. Two-neutron knockout from intermediate-energy beams in inverse kinematics provides therefore an alternative spectroscopic tool, even though with a different selectivity with respect to two-neutron transfer \cite{tos04,tos06}. It has been described as a direct reaction \cite{yon02,baz03,yon06}. While two-nucleon transfer selects the relative s-state of the pair, correlations driven by the two-nucleon knockout mechanism are much less selective and the configurations where the nucleons are close in space are favored \cite{tos04}. 
By removing two nucleons of the deficient species, one can populate even more exotic nuclei with significant cross sections (few mbarn) \cite{yon06,aud13}. From the experimental point of view one can benefit from the increased luminosity allowed by the use of thick targets at higher beam energy, and of the higher detection efficiency due to the kinematical focusing of the reaction products at forward angles. \\
In this paper we report on the in-beam $\gamma$ spectroscopy of the neutron deficient isotopes $^{102}$Sn and $^{100}$Cd, populated by intermediate-energy two-neutron removal reactions. The same reaction on the stable $^{112}$Sn isotope has been measured as a reference. New states are populated in the two isotopes under study and their nature is discussed based on two-neutron knockout cross sections.

\section{\label{sec:level2} Experimental method and results}
The experiment was performed at the Radioactive Isotope Beam Factory (RIBF) operated by the RIKEN Nishina Center and the Center for Nuclear Study (CNS) of the University of Tokyo. 
The $^{104}$Sn and $^{102}$Cd radioactive beams were produced with an intensity of 350 pps and 120 pps, respectively, via fragmentation of 6 pnA $^{124}$Xe primary beam at 345 MeV/u on a 555 mg.cm$^{-2}$ Be target and separated through the BigRIPS separator \cite{kub12}. Secondary beam energies at mid CH$_2$ target were 150 MeV/u for $^{104}$Sn, 144 MeV/u for $^{102}$Cd, and 170 MeV/u for the reference stable beam $^{112}$Sn. The purities of the main isotopes of interest were as follows: $^{104}$Sn (20\%), $^{102}$Cd (8\%) for the $^{104}$Sn setting, $^{112}$Sn (75\%) for the $^{112}$Sn setting. 
Two secondary targets were successively used for the measurement: a CH$_2$ target of 192(4) mg.cm$^{-2}$ thickness and a C target of 370(7) mg.cm$^{-2}$ thickness. The target thicknesses were determined both by weighting and by measuring the magnetic-rigidity deviation of the beam in the ZeroDegree spectrometer after the secondary target.
The cross section on hydrogen was extracted from the CH$_2$ data after subtraction of the normalized carbon component. Inclusive cross sections for two-nucleon removal from C and H were found to be rather similar \cite{aud13}, as shown in Tab. \ref{t1}.
The setup consisted of the DALI2 array composed of 186 NaI scintillators \cite{tak14} for gamma-ray detection and the ZeroDegree spectrometer for downstream particle identification \cite{aud13}. A mass resolution of $\sigma$ $\sim$ 0.001 was achieved, allowing for an unambiguous isotopic identification. 
The scintillators of the DALI2 array were calibrated in energy with $^{137}$Cs, $^{88}$Y and $^{60}$Co sources, with gamma emission ranging between 661 and 1836 keV. The efficiency of the DALI2 array, covering angles between 19 and 150 deg., was 14\% at 1.33 MeV. This value was in agreement within 6\% with the Geant4 simulation \cite{geant} and was rather independent of the angular distribution of the emitted gamma radiation thanks to the large solid angle coverage of the DALI2 array. The FWHM intrinsic energy resolution (in keV) evaluated from Cs, Co and Y sources scaled as 2.4$\sqrt{E}$, where E is given in keV, consistent with \cite{tak14}.

\begin{figure}[h!]
\begin{center}
     \begin{tikzpicture}
     \node at (1,1) {\includegraphics[width=0.52\textwidth]{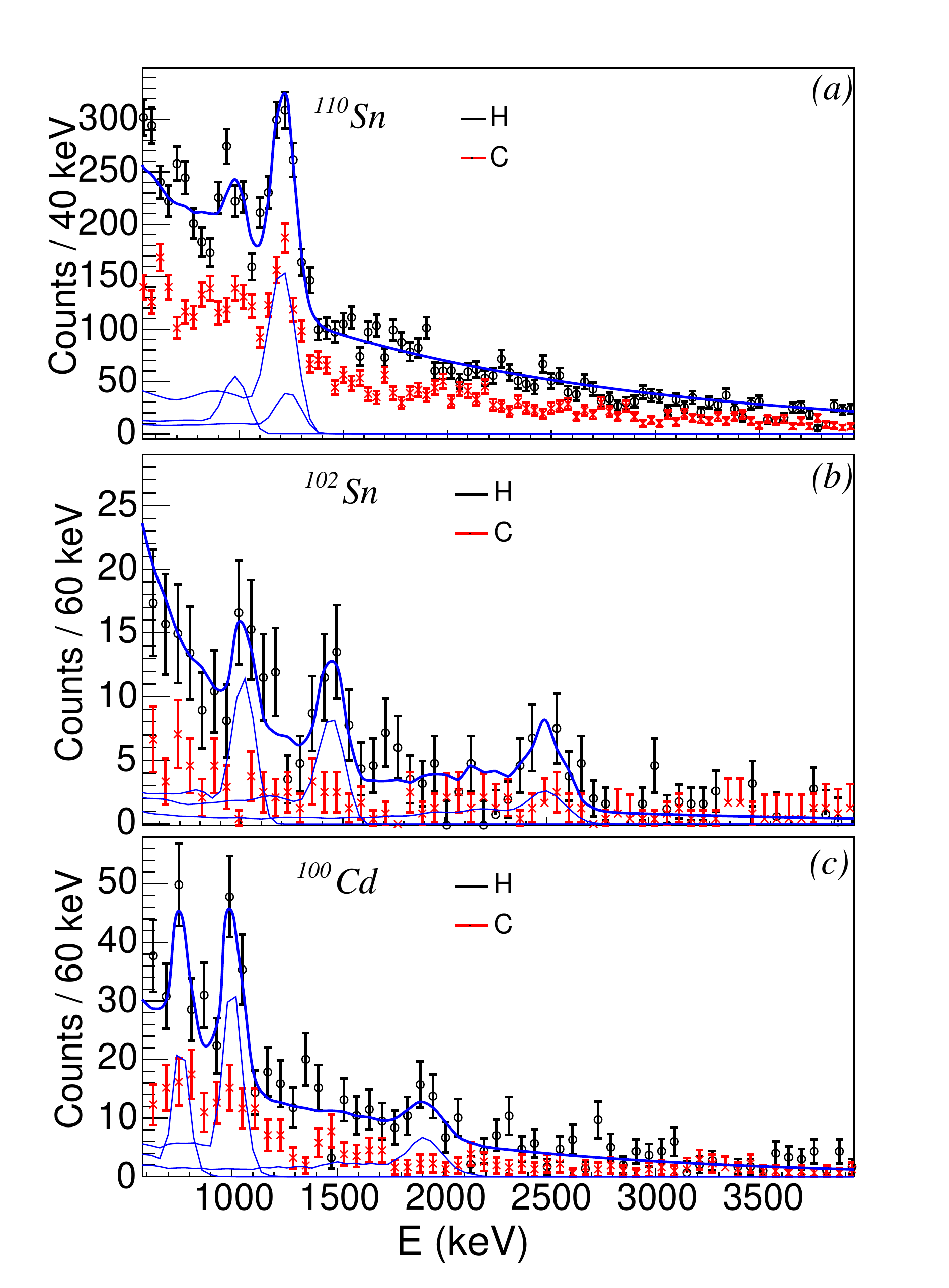}};
     \node at (3.,5.3) {\includegraphics[width=0.18\textwidth]{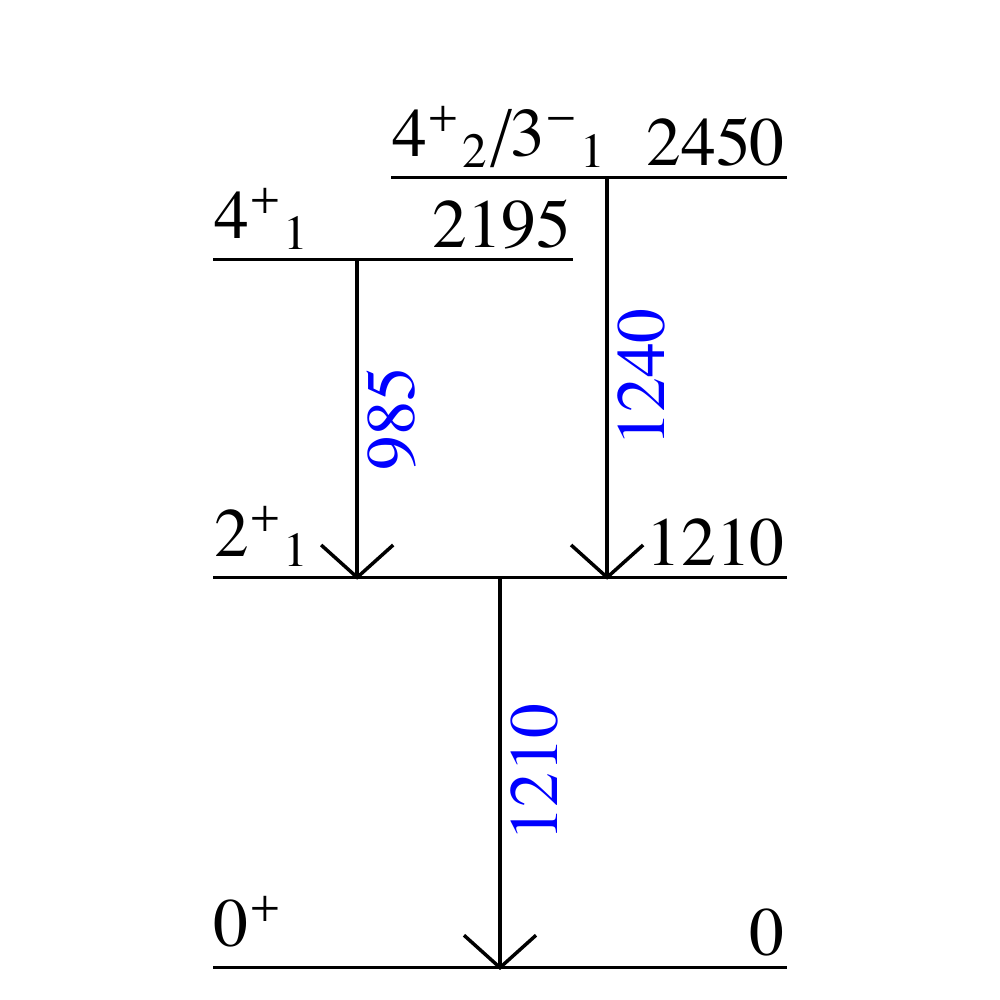}};
     \node at (3.,1.5) {\includegraphics[width=0.17\textwidth]{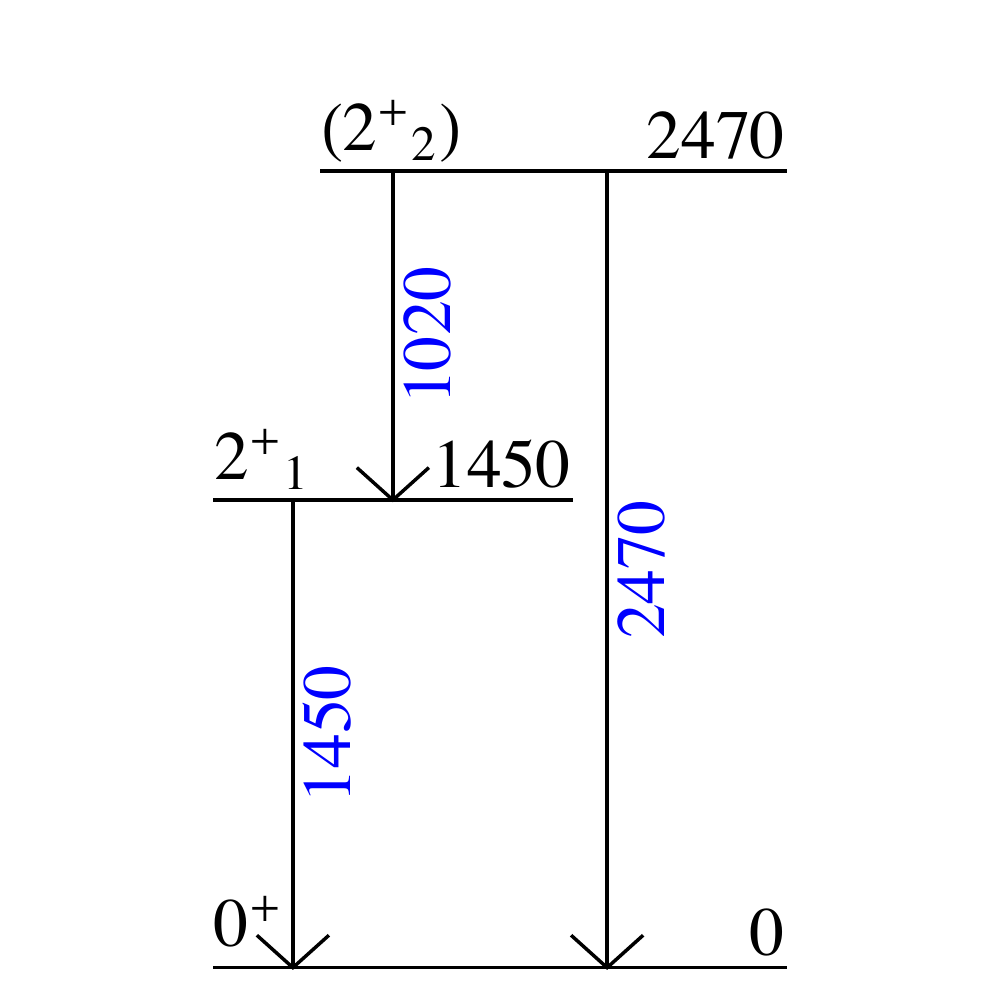}};
      \node at (3.,-1.5) {\includegraphics[width=0.18\textwidth]{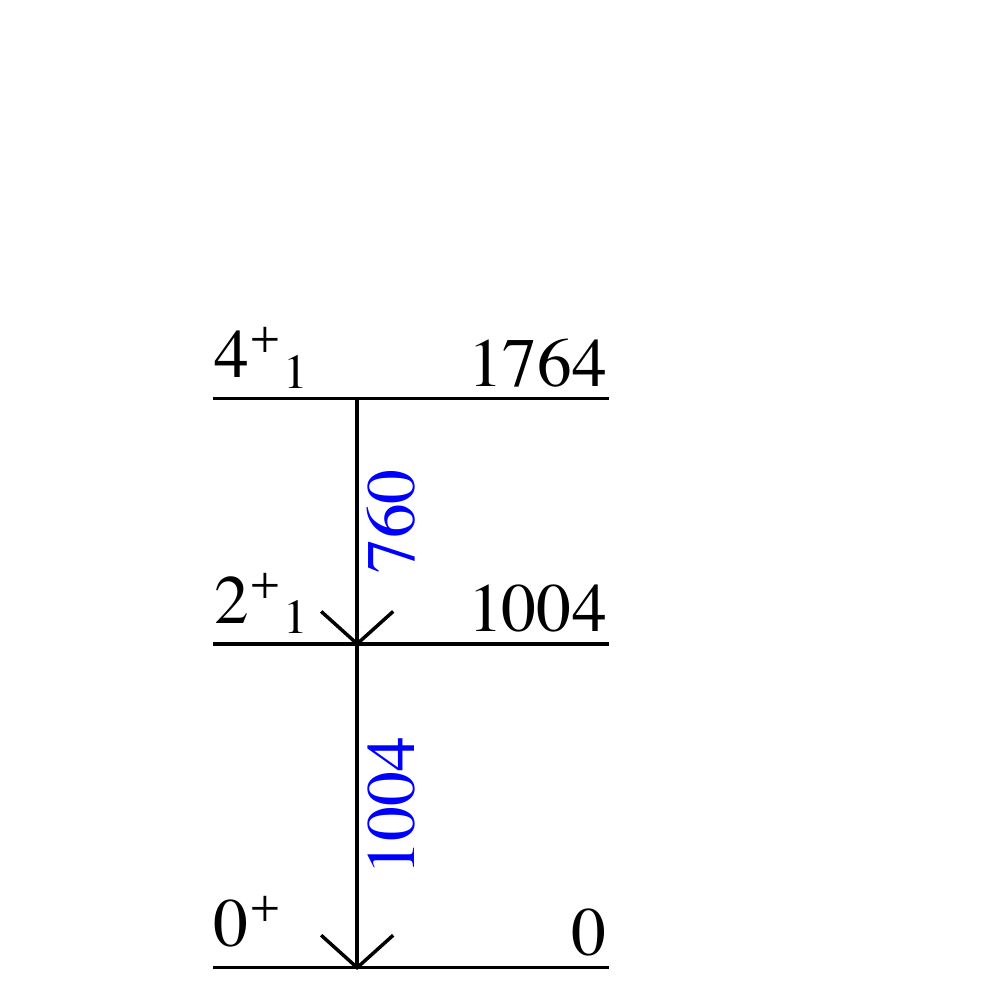}};
       \end{tikzpicture}
 \caption{\label{fall} (Color online) Gamma spectra measured after two-neutron removal from C (red crosses) and CH$_2$ targets (after C subtraction, black dots), fitted with the sum of the detector response functions and two exponential backgrounds (blue line). The projectile-like reaction products are (a) $^{110}$Sn, (b) $^{102}$Sn, (c) $^{100}$Cd. The corresponding level schemes are shown, with energies in keV. }
\end{center}
\end{figure}

\begin{figure}[h]
\begin{center}
      \includegraphics[width=8.7cm,trim=0cm 9.5cm 0cm 10cm]{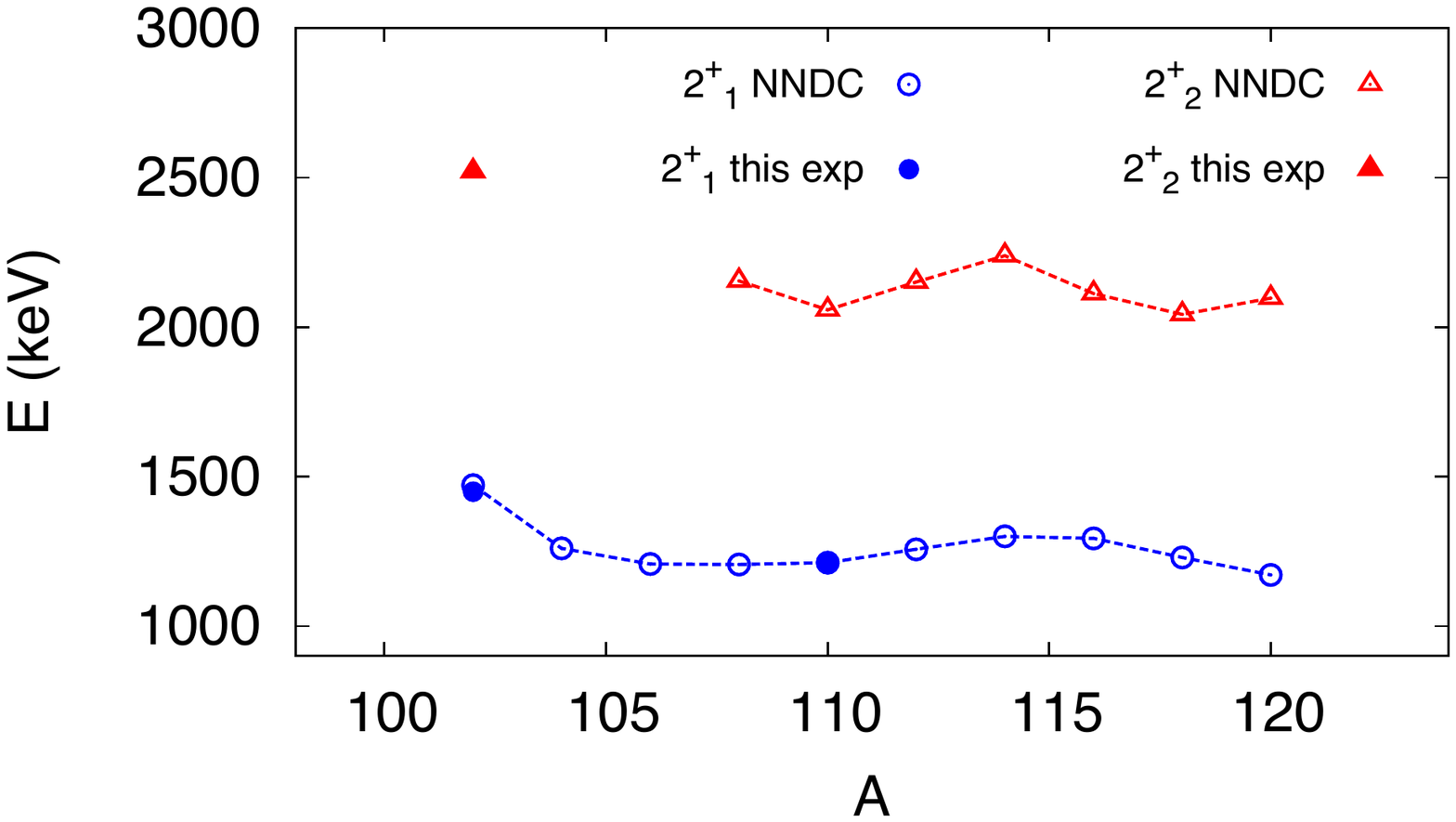}
    \caption{\label{f1} (Color online) Systematics of 2$^+_1$ and 2$^+_2$ states in tin isotopes from $^{102}$Sn to $^{120}$Sn from Ref. \cite{nndc} (NNDC) and this experiment.} 
\end{center}
\end{figure}

In order to extract the cross sections to the excited states observed in this experiment, the spectrum was fitted with the sum of a double-exponential background and the simulated response functions of the DALI2 array to the gamma-transitions mentioned above. The background originated mainly from atomic processes (bremsstrahlung in the target) at low energy (E$_\gamma$ $<$ 500 keV), from Compton scattering and target breakup at higher gamma energy. It has been modeled with the sum of two exponentials, as done previously in similar analyses \cite{dor13b}.  An isotropic distribution in the center-of-mass frame of the emitting nucleus has been assumed in the simulation. A relative variation of 10\% in the efficiency is observed if an anisotropic distribution as the one in Ref. \cite{tak14} is assumed and is taken into account in the uncertainties on the cross section. The acceptance of the ZeroDegree spectrometer has been evaluated with a dedicated run with the spectrometer centered on the magnetic rigidity of the beam. The ratio between the transmitted beam and the incident beam yields the particle-identification efficiency (including the acceptance of the spectrometer, target effect, beam-line detectors efficiencies) and spans between 50\% and 65\% for the settings of interest and for the fully-stripped ions. The acceptance is expected to be the same for the two-neutron knockout fragments, once the spectrometer is set on their magnetic rigidity as during the rest of the measurement. The x-position distribution of two-neutron knockout fragments at the dispersive focal plane of ZeroDegree spectrometer shows that the acceptance does not induce any cut in the kinematics of the reaction.\\
Both spectra with and without add-back between adjacent crystals of the DALI2 array have been analyzed. Cross sections extracted in the two ways are found to be consistent within the experimental error bars, therefore add-back spectra are used to maximize statistics in the peaks.
The error on the cross section was obtained as the sum of the error issued from the $\chi^2$ minimization (which includes statistical error), the statistical error on the number of gamma counts, the uncertainties on the target thickness (2\%), the DALI2 efficiency (6\%), the acceptance of ZeroDegree spectrometer (3\%), and the background determination. This last source of error was fixed by varying the background amplitude within the error bars in the regions free of transitions, and propagating this variation on the cross section. Its magnitude is comparable or up to twice the sum of the statistical errors, depending on the energy of the gamma transition.\\
We discuss first our reference case, $^{110}$Sn. The spectroscopy of $^{110}$Sn is well known since it has been studied via several reaction probes including the $^{112}$Sn$(p,t)$ reaction \cite{gua06}. Fig. \ref{fall} (a) shows the spectrum of $^{110}$Sn produced via knockout on C (red) and on CH$_2$ targets, subtracted from C contaminations after normalization (black). Two peaks were observed at $\sim$ 900-1000 keV and at $\sim$ 1200 keV. The spectrum was fitted with the sum of the transitions corresponding to the gamma decay of the 2$^{+}_1$ to the ground state, and of the 4$^{+}_1$, 4$^{+}_2$/3$^{-}_1$ states to the 2$^{+}_1$ state \cite{gua06}. The two-neutron knockout reaction mechanism populates preferentially the same states as the two-neutron transfer reaction reported in Ref. \cite{gua06}. Cross sections to the observed states are summarized in Tab. \ref{t1}. \\ 
The cross section to the ground state can be extracted by subtracting the sum of cross sections to the observed excited states from the inclusive cross section. By doing so, a cross section to the ground state of 82.9(24) mb, and a ratio R=$\sigma_{gs}$/$\sigma_{inclusive}$=66(7)\% were obtained. 
In this experiment we do not observe the decay of the known 2$^{+}_2$ state at 2121 keV via the 909 keV and 2121 keV transitions \cite{nndc}. Note that one cannot exclude unobserved feeding from this or other higher-lying excited states, which is difficult to evaluate. We can put an upper limit of 4 mb to such feeding by evaluating the cross section corresponding to the number of counts that would be necessary to identify a transition at 2121 keV.\\
We discuss now the spectroscopy of the N=52 isotones $^{102}$Sn, $^{100}$Cd populated via two-neutron removal from $^{104}$Sn and $^{102}$Cd, respectively. For $^{102}$Sn (Fig. \ref{fall} (b)), a transition at 1450(20) keV compatible with the known 2$^+_1$~$\rightarrow$~0$^+_1$ transition at 1472 keV was observed. A transition at 1020(20) was also observed, which for consistency with the total reaction cross section must feed one of the excited states. Statistics was too low to observe gamma-gamma coincidences. Nevertheless the 1020-keV transition, together with the 1450-keV one, sum up to 2470 keV, consistent with the energy of the third observed transition at 2470(60) keV. This level energy does not match known ones in $^{102}$Sn \cite{lip98}. Therefore, we concluded that this state decays to the 2$^{+}_1$ state and to the ground state with a branching ratio of 30(5)\% and 70(8)\%, respectively. Based on that, we propose a 2$^{+}_2$ spin-parity assignment. 
Fig. \ref{f1} shows the systematics of 2$^+_1$ and 2$^+_2$ state energies in the tin isotopes. The 2$^+_2$ in $^{102}$Sn is about 200 keV higher than the last observed one in $^{108}$Sn. A similar shift going from $^{108}$Sn to $^{102}$Sn is observed for the 2$^+_1$ state, and could therefore be expected also for the 2$^+_2$ state, corroborating our assignment. 
The existence of the new transitions at 1020(20) and 2470(60) keV was assessed based on the Bayesian Information Criterion (BIC) \cite{bic}. According to this criterion, the model with the highest probability is the one that minimizes the BIC = -2 $\times$ (log maximized likelihood) + (log N) $\times$ (number of parameters), where N is the number of bins and the number of parameters is the number of functions used to fit the spectrum.
According to this criterion, the evidence against higher BIC of the fit including these transitions is 21 and 12, respectively, that corresponds to a very strong evidence.  \\
In this experiment we could not observe the transition at 497 keV corresponding to the decay of the known 4$^+_1$ to the 2$^+_1$ state. 
This can be due to the fact that the energy spectrum below 600 keV is dominated by an exponential background associated to target breakup, Compton scattering from higher-energy transitions, and Bremsstrahlung.
Even though we cannot rule it out completely, a significant feeding of the 2$^+_1$ state by the 4$^+_1$ is unlikely since it will correspond to an even smaller direct population of the 2$^+_1$ state.

For $^{100}$Cd (Fig. \ref{fall} (c)), the known 2$^{+}_1$~$\rightarrow$~0$^{+}_1$ and 4$^{+}_1$~$\rightarrow$~2$^{+}_1$ transitions at 1004(15) keV and 760(15) keV, respectively, were observed. Furthermore, accumulations of statistics at higher energy, around 1340 keV and 1930 keV, were observed. We applied the same BIC criterion, and only the second peak at 1930(20) keV survives with an evidence against higher BIC of 2.5. A natural candidate for this state would be a 2$^+_2$ state, which is unknown in $^{100}$Cd, but statistics are too low for gamma-gamma coincidences and a definite assignment for this transition cannot be provided. The possibility that this transition feeds the 2$^{+}_1$ and 4$^{+}_1$ states is taken into account in the error bars as systematic uncertainties. Cross sections are detailed in Tab. \ref{t1}. We remark that if this transition decays exclusively to 2$^{+}_1$ and 4$^{+}_1$ states, this will imply no direct population of these states, which seems rather unlikely.\\ 
All measured cross sections are summarized in Tab. \ref{t1}. Both in the case of $^{104}$Sn(H,X)$^{102}$Sn and $^{102}$Cd(H,X)$^{100}$Cd, a ratio R=$\sigma_{gs}$/$\sigma_{inclusive}$ of 28(9)\% and 30(3)\% was measured, very different from the one observed for $^{112}$Sn(H,X)$^{110}$Sn, 66(7)\%. We have no quantitative explanation for this effect at this stage, though it recalls the enhancement of the two-neutron transfer cross section to the low-lying states associated to the evolution of the ground state from superfluid to non-superfluid regime approaching the shell closure. Again, the cross section to the ground state extracted here may be affected by unobserved transitions. Such eventuality would imply an even smaller cross section to the ground state, amplifying the effect described above. \\

\begin{table}[h!]
\centering
\begin{tabular}{ccccc}
\hline
\hline
\multirow{2}{*}{Nucleus}& \multirow{2}{*}{J$^\pi$}  & E$_{ex}$&   E$_\gamma$& $\sigma_{exp}$ \\
& &(keV)&(keV)&(mb) \\
\hline
 
 \multirow{5}{*}{$^{110}$Sn}&2$^{+}_1$&1210&1210*&17.3(26)\\
&4$^{+}_1$&2195&985*&10.5(8)\\
&4$^{+}_2$, 3$^{-}_1$&2450&1240*&8.7(12)\\
&&&inclusive (H)&107(7) \\
&&&inclusive (C)&98(4) \\
\hline
\multirow{4}{*}{$^{102}$Sn}&2$^{+}_1$ & 1450(20) & 1450(20) & 0.5(2)\\
&(2$^{+}_2)$ & 2470(30) & 2470(60), 1020(20) & 1.4(9)\\
&&&inclusive (H)&2.6(3) \\
&&&inclusive (C)&2.1(1) \\
\hline
\multirow{5}{*}{$^{100}$Cd}&2$^{+}_1$&1004(15)&1004(15)&3(1)$^{0}_{-3}$\\
&4$^{+}_1$&1764(20)&760(15)&2.7(6)$^{0}_{-3}$\\
&&1930+X&1930(20)&3(3)\\
&&&inclusive (H)& 11.7(6)\\
&&&inclusive (C)& 8.9(3)\\
\hline
\hline
\end{tabular}
\caption{\label{t1} Two-nucleon removal cross sections (exclusive for H, inclusive for H and C) at about 150 MeV/nucleon (see text for details) to the states observed in the A-2 nuclei $^{110}$Sn, $^{102}$Sn and $^{100}$Cd. Transition energies marked with an asterisk are taken from Ref. \cite{nndc}, while other energies are extracted from our measurement. Only systematic error on the cross section due to observed feeding is taken into account.}
\end{table}

\section{\label{sec:level3} Interpretation}
An interpretation of the enhanced cross section to the  2$^{+}_2$ state in $^{102}$Sn with respect to the 2$^{+}_1$ state can be found in a modification of the structure of the Sn isotopes approaching N=50, that can be pinned down via a shell model calculation.

\begin{figure}[h]
\begin{center}
      \includegraphics[trim={0 8cm 0 8cm},clip,width=0.5\textwidth]{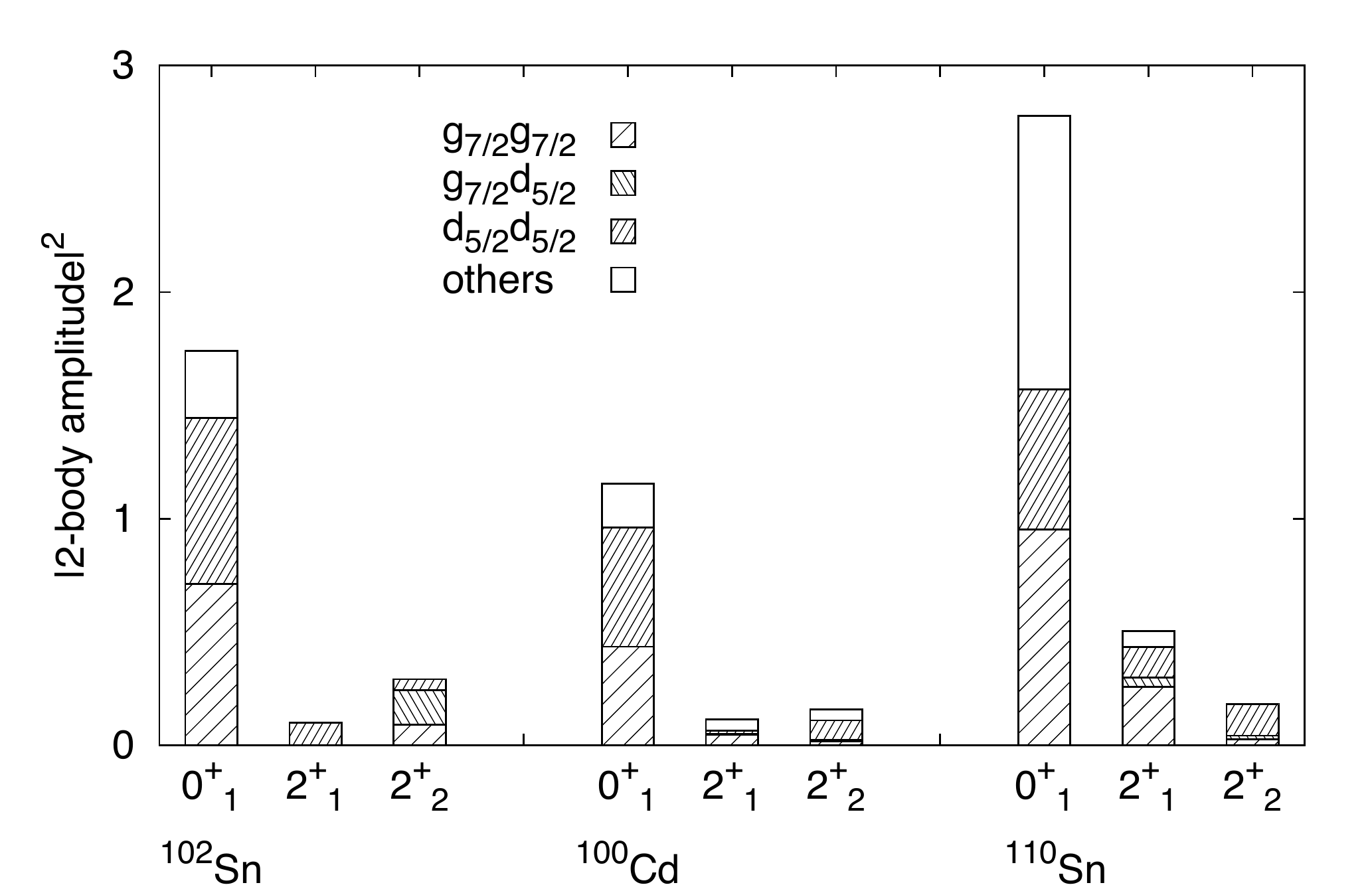}

    \caption{\label{overlap} (Color online) Two-body overlaps between A and A-2 nuclei from shell-model calculations.  Only main contributions (probability$>$0.1) from different shell-model configurations are shown.} 
\end{center}
\end{figure}

A shell model calculation was performed for the isotopes of interest in the neutron g$_{7/2}$d$_{5/2}$d$_{3/2}$s$_{1/2}$h$_{11/2}$ model space for $^{102, 104, 110, 112}$Sn using the interaction of Ref. \cite{ban05}. In the case of $^{100, 102}$Cd the proton g$_{9/2}$ orbital was added to the model space. 
The resulting excitation energy spectra of $^{110,102}$Sn and $^{100}$Cd are in good agreement with experimental results as far as 2$^+$ states are concerned. 
The agreement between shell-model calculation and experimental data, as far as 2$^+$ states are concerned, is as follows: in $^{110}$Sn the 2$^+_1$ state energy is overestimated by 84 keV; in $^{102}$Sn the 2$^+_1$ state energy is underestimated by 64 keV and the 2$^+_2$ state energy overestimated by 543 keV; in $^{100}$Cd the 2$^+_1$ state energy is underestimated by 153 keV. The ground state of $^{104}$Sn appears to be dominated by the (g$_{7/2}$)$^2_{0^+}$(d$_{5/2}$)$^2_{0^+}$ configuration (86\%) as well as the $2^+_2$ state in $^{102}$Sn by the g$_{7/2}$d$_{5/2}$ configuration (39\%).\\
To compare the evolution of two-neutron knockout cross sections, assuming a direct two-nucleon removal process, the most suitable quantities are two-body overlaps between A nuclei ($^{112, 104}$Sn(gs), $^{102}$Cd(gs)) and different low-lying states of A-2 nuclei ($^{110, 102}$Sn and $^{100}$Cd). Contributions from different shell-model configurations to two-body amplitudes are juxtaposed in Fig. \ref{overlap}, to give an idea of the total strength. 

At first glance, one can see that the largest overlaps occur when the final state of the A-2 nuclei is the ground state. The effect is particularly strong for the most exotic species, $^{102}$Sn and $^{100}$Cd, at difference with the experimental data signaling a weaker population of the ground state with respect to the excited states in the case of $^{102}$Sn and $^{100}$Cd, where only 28(9)°\% and 30(3)\% of the cross section goes to the ground state, respectively.\\
If we focus on the excited states of $^{102}$Sn, an enhancement of the two-neutron transfer probability to the 2$^{+}_2$ state of $^{102}$Sn is found and associated to the contribution from the transfer of a d$_{5/2}$g$_{7/2}$ pair. This prediction is in qualitative agreement with our experimental findings that correspond to $\sim$ 3 times larger two-neutron knockout cross section to the 2$^+_2$ state with respect to the 2$^+_1$ state. Such enhancement is not observed in $^{110}$Sn and $^{100}$Cd, in agreement with the fact that we could not observe signatures of population of the 2$^{+}_2$ state in our data.

These remarks cannot be made more quantitative until a microscopic description of proton-induced two-nucleons knockout cross sections at intermediate energies will be available. Several theoretical developments have been undertaken in recent years to calculate one-nucleon knockout cross sections \cite{ber13, cre14, oga15}. Theoretical models for two-nucleons knockout cross section on a nuclear target became recently available \cite{tos04, tos06}, whereas so far no theoretical model is able to predict two-nucleon removal cross sections on hydrogen.
\\

\section{\label{sec:level1} Conclusions and perspectives}
We presented new transitions measured in $^{102}$Sn and $^{100}$Cd populated via two-neutron removal from a proton target at about 150 MeV/nucleon beam energy. The transition with an energy of 2470(60) keV observed in $^{102}$Sn was assigned as the decay from the 2$^+_2$ state to the ground state. \\
We interpreted the enhancement of the two-neutron knockout cross section to this 2$^+_2$ state in $^{102}$Sn in terms of structural overlaps. The fact that the increased cross section to the 2$^+_2$ state in $^{102}$Sn corresponds to a larger structure overlap with respect to the 2$^+_1$ state is consistent with the assumption that proton-induced two-neutron removal can be interpreted as a direct process.
A complete interpretation of these cross sections would require a microscopic approach to predict the two-neutron removal cross sections from hydrogen which currently is not available. \\
Since we used a CH$_2$ target to extract the cross section on H we measured also cross sections on C for background subtraction. 
Interestingly, we remark that the inclusive cross sections on C and H are rather similar, as discussed in detail in \cite{aud13}. A systematic comparison of the relative population of bound states from heavy-ion and proton induced one- and two-nucleon knockout with high statistics in loosely bound nuclei may provide new insight into the reaction mechanism. \\
The spectroscopy of $^{100}$Sn and the measurement of B(E2) in $^{102}$Sn are two key experiments to characterize the shell closure at Z, N=50 and further test the robustness of magic numbers across the nuclear landscape. The present results may be used to evaluate the feasibility of measuring the gamma spectroscopy of $^{100}$Sn populated via one- and two-neutron removal, if the excited states of this nucleus decay via gamma emission. In fact, despite the low proton and alpha separation energy, the 2$^+_1$ state is expected to decay essentially via gamma emission due to the large Coulomb barrier. Indeed one should be careful in doing such extrapolations since structural changes are expected close to $^{100}$Sn. A high 2$^{+}_1$ energy around 4-5 MeV is expected for this doubly magic nucleus (Refs. \cite{fae13,mor18}). Experimentally, the use of a thick pure liquid hydrogen target presents several advantages: cleaner reaction mechanism, maximum ratio of luminosity over energy loss of the fragment, no need of carbon background subtraction as in the present experiment. Based on our measured cross-sections for $^{102}$Sn and the augmented primary beam intensities now available at the RIBF, we assess the spectroscopy of $^{100}$Sn populated via proton-induced two-neutron removal feasible within acceptable amounts of beam time.

\begin{acknowledgments}
This work has been supported by the European Research Council through the ERC Starting Grant No. MINOS-258567 and by the ENSAR European FP7 Project No. 262010. Y. Tanimura acknowledges the financial support from the Graduate Program on Physics for the Universe of Tohoku University. A. Jungclaus acknowledges support from the Spanish Ministerio de Econom\'ia y Competitividad under contract FPA2014-57196-C5-4-P. D. Sohler acknowledges support from the European Regional Development Fund (Contract No. GINOP-2.3.3-15-2016-00034). A. Corsi acknowledges S. P\'eru and M. Martini for interesting discussions, and N. Paul for careful proofreading.
\end{acknowledgments}

\bibliography{biblio}

\end{document}